\documentclass{sig-alternate}
\usepackage{algpseudocode}
\usepackage{algorithm}
\newcommand{\mdeg}[1]{\left< #1 \right>}

\begin{document}
%

\title{Recommendation systems in the scope of opinion formation: a model}
%
%
%
%
%
\numberofauthors{2} 
%
\author{
%
%
\alignauthor
Marcel Blattner\\
       \affaddr{Laboratory for Web Science}\\
       \affaddr{University of Applied Sciences FFHS}\\
       \affaddr{Regensdorf, Switzerland}\\
       \email{marcel.blattner@ffhs.ch}
\alignauthor
Matus Medo\\
       \affaddr{Physics Department}\\
       \affaddr{University of Fribourg}\\
       \affaddr{Fribourg, Switzerland}\\
       \email{matus.medo@unifr.ch}
       }

\date{30 July 1999}
\maketitle

\begin{abstract}
Aggregated data in real world recommender applications often feature fat-tailed
distributions of the number of times individual items have been rated or favored. We propose a model to simulate such data. The model is mainly based on social interactions and opinion formation taking place on a complex network with a given topology. A threshold mechanism is used to
govern the decision making process that determines whether a user is or is not interested
in an item. We demonstrate the validity of the model by fitting attendance distributions 
from different real data sets. The model is mathematically analyzed by investigating its master equation. Our approach provides an attempt to understand recommender system's data as a social process. The model can serve as a starting point to generate artificial data sets useful for testing and evaluating recommender systems.
\end{abstract}

\category{H.1.m}{Information Systems}{Miscellaneous}
\terms{Experimentation, Theory}
\keywords{recommender systems, opinion formation, complex networks}
\section{Introduction}
This is the information age. We are witnessing
information production and consumption in a speed never
seen before. The WEB2.0 paradigm enables consumers and producers to
exchange data in a collaborative way benefiting both parties. However,
one of the key challenges in our digitally-driven society is information
overload \cite{berg10}. We have the 'pain of choice'.
Recommendation systems represent a possible solution to this problem. They
have emerged as a research area on its own in the 90s
\cite{Res97,gold92,KoMi97,GoSc99,BrHe98}. The interest in recommendation systems increased steadily
in recent years, and attracted researchers from different fields \cite{recsyshand}. The success of highly rated Internet sites as Amazon, Netflix, YouTube, Yahoo, Last.fm and others is to a large extent based on their recommender engines. Corresponding
applications recommend everything from CD/DVD's, movies, jokes, books,
web sites to more complex items such as financial services. 

The most popular techniques related to recommendation systems are
collaborative filtering \cite{BiPa98,GoRo01,BrHe98,HeKo04,KoMi97,GoSc99,ReIa94,SaKa01} and
content-based filtering \cite{ClGo99,Pa07,MeMo02,bal97,lin03}.
In addition, researchers developed alternative methods inspired by
fields as diverse as machine learning, graph theory, and
physics \cite{fou07a,fou07b,mir03,zhou07,ZhBlYk07,Blattner10,web01,zha02}.
Furthermore, recommendation systems have been investigated in
connection with trust \cite{and08,don05,wal08,mas04,mas04b} and
personalized web search \cite{bir05,bru07,sug04}, which constitutes the
new research frontier in search engines.

However, there are still many open challenges in the research field of recommendation
systems \cite{AdTu05,guy10,jan09,gey10,HeKo04,recsyshand,dra10}. One key question is
connected to the understanding of the user rating mechanism.
We build on a well documented influence of social interactions with peers on the decision to vote, favor, or even purchase an item \cite{rich88,kim07}.
We propose a model inspired by opinion formation taking place on a complex network with a predefined topology. Our model is able to generate data observed in real world recommender systems. Despite its simplicity, the model is flexible enough to generate a wide range of different patterns. We
mathematically analyze the model using a mean field approach to the full Master
Equation. Our approach provides an understanding of the data in recommender systems as
a product of social processes. The model can serve as a data generator which is
valuable for testing and evaluation purposes for recommender systems.

The rest of the paper is organized as follows. The model is outlined in
Sec.~(\ref{sec:model}). Methods, data set descriptions, and validation procedures are in Sec.~(\ref{sec:methods}). Results are presented in Sec.~(\ref{sec:results}).
Discussion and an outlook for future research directions are in Sec.~(\ref{sec:discussion}).

\section{Model}
\label{sec:model}

\subsection{Motivation}
Our daily decisions are heavily influenced by various information
channels: advertisement, broadcastings, social interactions, and
many others. Social ties (word-of-mouth) play a pivotal role in consumers buying
decisions \cite{rich88,kim07}. It was demonstrated by many
researchers that personal communication and informal information
exchange not only influence purchase decisions and opinions, but shape our
expectations of a product or service \cite{wh54,arndt67,and03}.
On the other hand, it was shown \cite{henn04}, that social benefits are a
major motivation to participate on opinion platforms. 
If somebody is influenced by recommendations on an opinion platform like MovieLens or Amazon,
social interactions and word-of-mouth in general are additional
forces governing the decision making process to purchase or even to
rate an object in a particular way \cite{mas09}. 

Our model is formulated within an opinion formation framework
where social ties play a major role. We shall discuss the following main ingredients of our model:
\begin{itemize}
\item Influence-Network (IN)
\item Intrinsic-Item-Anticipation (IIA)
\item Influence-Dynamics (ID)
\end{itemize}

\paragraph{Influence Network} We call the network where context-relevant
information exchange takes place an Influence-Network
(IN). Nodes of the IN are people and connections between nodes indicate
the influence among them. Note that we put no constraints on the
nature of how these connections are realized. They may be purely 
virtual (over the Internet) or based on physical meetings.
We emphasize that INs are domain dependent, i.e., for a given community of users, the Influence Network
concerning books may differ greatly (in topology, number of ties, tie strength, etc.) from that concerning
another subject such as food or movies. Indeed, one person's
opinion leaders (relevant peers) concerning books may be very different from those for food or other
subjects. In this scope, we see the INs as domain-restricted views on
social networks. It is thus reasonable to assume that Influence Networks are similar to social interaction networks which often exhibit a scale-free topology \cite{bar08}. However, our model is not restricted to a particular network structure.

\paragraph{Intrinsic-Item-Anticipation} Suppose a new product is launched on the market. Advertisement, marketing campaigns, and other efforts to attract customers predate
the launching process and continue after the product started to
spread on the market. These efforts influence product-dependent
customer anticipation. It is clear that the resulting anticipation is
a complex combination of many different components including intrinsic
product quality and possibly also suggestions from recommendation systems.

In our model we call the above-described anticipation
Intrinsic-Item-Anticipation (IIA) and measure it by a single number.
It is based on many external sources, except for the influence generated by social
interactions. It is the opinion on something taken by individuals, before they
start to discuss the subject with their peers. Furthermore, we assume
that an individual will invest resources (time/money) into an object
only, if the Intrinsic-Item-Anticipation is above a particular
threshold, which we call Critical-Anticipation-Threshold.  

\paragraph{Influence-Dynamics} The Influence-Dynamics describes
how individuals' Intrinsic-Item-Anticipations are altered by information exchange via 
the connections of the corresponding Influence-Network.
From our model's point of view this means the following: an individual's
IIA for a particular item $i$ may be shifted
due to social interactions with directly connected peers (these interactions thus take place on the corresponding IN), who already experienced the product or service in question. This process
can shift the Intrinsic-Item-Anticipation of an individual who did not yet experience
product/object $i$ closer to or beyond the critical-anticipation-threshold.

We now summarize the basic ingredients of our model. An individual user's opinions on
objects are assembled in two consecutive stages: i) opinion making based on different external sources, including suggestions by recommendation systems and ii) opinion making based on social
interactions in the Influence-Network. The second process may shift the opinions generated
by the first process.

\subsection{Mathematical formulation of the model}
\begin{figure}
\centering
\includegraphics[scale=0.35]{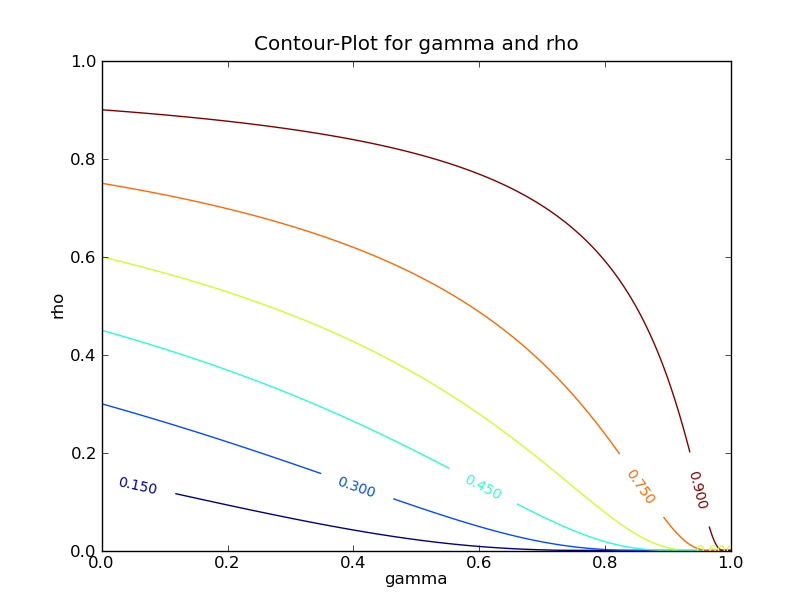}
\caption{Contour plot for $\gamma$ and $\rho=\Theta_{j}/k_{j}$.
Numbers inside the plot quantify the shift in the IAA as a function of $\gamma$ and $\rho$.}
\label{fig:conplot1}
\end{figure}

In this section we firstly describe how individuals' Intrinsic-Item-Anticipations may change due to social interactions taking place on a particular Influence-Network. Secondly, we introduce dynamical processes governing the opinion propagation.
\paragraph{IIA shift}
We model a possible shift in the IIA as:
\begin{equation}
\label{equ:iia1}
\hat{f}_{i j} = f_{i j} +  \left[\frac{\Theta_{j}}{k_{j}}\right]^{(1-\gamma)}.
\end{equation}
where $\hat{f}_{ij}$ is the shifted Intrinsic-Item-Anticipation of
individual $j$ for object $i$, $f_{ij}$ is the unbiased
IIA, $\Theta_{j}$ is the number of
$j$'s neighbors, who already experienced and liked item $i$, $k_{j}$ denotes
the total number of $j$'s neighbors in the corresponding IN, and $\gamma \in (0,1)$ quantifies trust of individuals to their peers. An individual $j$ will consume, purchase, or positively rate an item $i$ only if
\begin{equation}
\label{equ:iia_cond}
\hat{f}_{ij} \ge \Delta.
\end{equation}
We identify $\Delta$ as the Critical-Anticipation-Threshold. 
Values of $f_{ij}$ are drawn from a probability distribution $f_{i}$. Since the IIA for each individual is an aggregate of many different and largely independent contributions,
we assume that $f_{i}$ is normally distributed, 
$f_{i} \in \mathcal{N}(\mu_{i},\sigma)$. (Unless stated otherwise.)
To mimic different item anticipations for different objects $i$, we draw the mean $\mu_{i}$
from a uniform distribution $U(-\epsilon,\epsilon)$. We maintain $\mu_{i}$, $\epsilon$, and
$\sigma$, so that $f_{i}$ is roughly bounded by $(-1,1)$, i.e.,
$-1 \le \mu - 3 \sigma < \mu + 3 \sigma \le 1$. Note that $\hat{f}_{ij}$
can exceed these boundaries after a shift of the corresponding IIA occurs.
The second term on the right hand side of Eq.~(\ref{equ:iia1}) is the
influence of $j$'s neighborhood weighted by trust $\gamma$.
To better understand the interplay between $\gamma$ and the density of attending users in the neighborhood of user $i$, $\rho:=\Theta_{j}/k_{j}$, we refer to Fig.~\ref{fig:conplot1}. Trust $\gamma\approx 1$ causes a big shift on the IIA's even for $\rho \approx 0$. On the
other hand, $\gamma \approx 0$ needs high $\rho$ to yield a significant IAA shift. These properties are understood as follows: people trusting strongly in their
peers need only few positive opinions to be convinced, whereas people
trusting less in their social environment need considerable more
signals to be influenced.
\paragraph{Influence-Dynamics}
The Influence-Dynamics proceeds as follows. Firstly, we draw an Influence-Network IN$(\mathcal{P})$ with a fixed network topology (power-law, Erd\H{o}s-R\'enyi, or another).
$\mathcal{P}$ refers to a set of appropriate parameters for the Influence-Network in question (like network type, number of nodes, etc.). The network's topology is not affected by the dynamical processes (opinion propagation) taking place on it. We justify this static scenario by assuming that the time scale of the topology change is much longer then the time scale \footnote{The term time scale denotes a dimensionless quantity and specifies the devisions of time. A shorter time scale means a faster spreading of opinions in the network.} of opinion spreading in the network.
Each node in the Influence-Network corresponds to an individual. For each
individual $j$ we draw an unbiased Intrinsic-Item-Anticipation $f_{ij}$
from the predefined probability distribution $f_{i}$.  
At each time step, every individual is in one of the following states: 
$\{S,A,D\}$. $S$ refers to a susceptible state and corresponds to the initial state for all nodes at $t=0$. $A$ refers to an attender state and corresponds to an individual with the property $\hat{f}_{ij} \ge \Delta$. $D$ refers to a denier state with the property $\hat{f}_{ij} < \Delta$ after an information exchange with his/her peers in the
Influence-Network happened. An individual in state $D$ or $A$ can not change
his/her state anymore. It is clear that an individual in state 
$A$ cannot back transform to the susceptible state $S$, since he/she did consume or
favor item $i$ and we do not account for multiple attendances in our model.
An individual in state $D$ was influenced but not convinced by his opinion leaders
(directed connected peers). We make the following assumption here: if individual $j$'s opinion
leaders are not able to convince individual $j$, meaning that individual's
$j$ Intrinsic Item Anticipation $\hat{f}_{ij}$ stays below the critical
threshold $\Delta$ after the influence process, then we assume that $j$'s opinion
not to attend object $i$ remains unchanged in the future.
Therefore we have the following possible transitions for each node in the influence network:
$j_{S} \rightarrow j_{A}$ or $j_{S} \rightarrow j_{D}$. Node states are updated asynchronously which is more realistic than synchronous updating, especially in social interaction models \cite{car08}.
The Influence-Dynamics is summarized in Algorithm \ref{recsysmodbas}.

\begin{algorithm}[!ht]
	\caption{RecSysMod algorithm. $\mathcal{P}$ contains the configuration parameter for the network. $\Delta$ is the Anticipation Threshold and $\gamma$ denotes the trust. $O \in \mathbb{N}$ is the number of objects to simulate. $G(N,E)$ is the network. $N$ is the set of nodes and $E$ is the set of edges.}\label{recsysmodbas}
\begin{algorithmic}[1]
\Procedure{RecSysMod\_I}{$\mathcal{P},\Delta,\gamma,O$}
\State $G(N,E)\gets$ GenNetwork($\mathcal{P}$) 
\ForAll{Objects in $O$}
\State generate distribution $f_{i}$ from $\mathcal{N}(\mu_{i},\sigma)$
\For{each node $j \in N$ in $G$} 
   \State draw $f_{ij}$ from $f_{i}$
   \If{$f_{ij} < \Delta$}
     \State $j_{state} \gets S$
   \Else
     \State $j_{state} \gets A$
   \EndIf 
\EndFor
\Repeat
\ForAll{$j$ with $j_{state}=S$ AND $ \Theta_{j} > 0$}
\State $\hat{f}_{ij} \gets  f_{ij} + 
\left[\frac{\Theta_{j}}{k_{j}}\right]^{(1-\gamma)}$
   \If{$\hat{f}_{ij} < \Delta$}
     \State $j_{state} \gets D$
   \Else
     \State $j_{state} \gets A$ 
   \EndIf 
\EndFor
\Until{$|\{j|j_{state}=S$ AND $\Theta_{j} > 0\}| = 0$}
\EndFor
\EndProcedure
\end{algorithmic}
\end{algorithm}

\paragraph{Master Equation}
We are now in the position to formulate the Master Equation for the dynamics. 
As already said before, two things can happen when a non-attender is connected to an attender:
a) he/she becomes an attender too, or b) he/she becomes a denier who will not attend/favor the item. For these two interaction types we formally write:
\begin{eqnarray}
S + A & \stackrel{\lambda}{\longrightarrow}& 2A \nonumber \\
S + A & \stackrel{\alpha}{\longrightarrow} & D + A
\end{eqnarray}
Here $\lambda$ denotes the probability that a susceptible node connected to an attender
becomes an attender too, and $\alpha$ is the probability that a susceptible node
attached to an attender becomes a denier. To take
into account the underlying network topology of the Influence Network it is common to
introduce compartments $k$ \cite{gle11}.
Let $N^{A}_{k}$ be the number of nodes in state $A$ with $k$
connections, $N^{S}_{k}$ the number of nodes in state $S$ with $k$ connections, and
$N^{D}_{k}$ the number of nodes in state $D$ with $k$ connections, respectively. Furthermore we define the corresponding densities: $a_{k}(t)=N^{A}_{k}/N_{k}$, $s_{k}(t)=N^{S}_{k}/N_{k}$ and
$d_{k}(t)=N^{D}_{k}/N_{k}$. $N_{k}$ is
the total number of nodes with $k$ connections in the network. Since every node from $N_k$ must be in one of the three states, $\forall t:\ a_{k}(t) + s_{k}(t) + d_{k}(t) = 1$.
A weighted sum over all $k$ compartments gives the total fraction of attenders at time $t$, $a(t) =
\sum_{k}P(k)a_{k}(t)$ where $P(k)$ is the degree distribution of the network (it also holds that $a(t)=N^A(t)/N$). The time dependence of our state variables $a_{k}(t),d_{k}(t),s_{k}(t)$ is
\begin{equation}
\left.\begin{aligned}
\dot{a}_{k}(t) &= \lambda k s_{k}(t) \Omega \\
\dot{d}_{k}(t) &= \alpha k s_{k}(t) \Omega \\
\dot{s}_{k}(t) &= -(\alpha + \lambda) k s_{k}(t) \Omega 
\end{aligned}\right\}
\label{eq:mas1}
\end{equation}
where $\Omega$ is the density of attenders in the neighborhood of susceptible node with $k$ connections averaged over $k$
\begin{equation}
\Omega = \sum_{k}P(k)(k-1)a_{k}/\mdeg{k}
\label{eq:omega}
\end{equation}
where $\mdeg{k}$ denotes the mean degree of the network. As outlined above, $\lambda$ is the probability that a node in state $S$ transforms to state $A$ if it is connected to a node in state $A$. This happens when $\hat{f}_{ij} > \Delta$. Therefore, we have $ \Delta_{-} < f_{ij} < \Delta$ where $\Delta_{-} = \Delta - (1/k)^{1-\gamma}$. From this we have
$\lambda = \int_{\Delta_{-}}^{\Delta}f(x)dx$, where $f(x)$ is the expectation distribution. Similarly we write for $\alpha = \int_{l}^{\Delta_{-}} f(x)dx$, where $l$ denotes the lower bound of the expectation distribution $f(x)$. A crude mean field approximation can be obtained by multiplying the right hand sides of Eq.~(\ref{eq:mas1}) with $P(k)$ and summing over $k$, which yields a set of differential equations
\begin{equation}
\left.\begin{aligned}
\dot{a}(t) &= \lambda \mdeg{k} s(t)a(t),\\
\dot{d}(t) &= \alpha \mdeg{k} s(t) a(t),\\
\dot{s}(t) &= -(\alpha + \lambda) \mdeg{k} s(t) a(t).
\end{aligned}
\right\}
\label{eq:mas2}
\end{equation}
which is later used to obtain analytical results for the attendance fraction $a(t)$.

\section{Methods}
\label{sec:methods}
We describe here our simulation procedures, datasets, experiments, and analytical methods.

\paragraph{Simulations} Our simulations employ Alg.~(\ref{recsysmodbas}). As outlined
in the model section, we do not change the network topology during the dynamical
processes. We experiment with two different network types, Erd\H{o}s-R\'enyi (ER),
and power law (PL) which are both generated by a so-called configuration model
\cite{new1}. ER and PL represent two fundamentally different classes of networks. The
former is characterized by a typical degree scale (mean degree of the network),
whereas the latter exhibits a fat-tailed degree distribution which is scale free. The
networks are random and have no degree correlations and no particular community
structure. To obtain representative results we stick to the following approach: we
fix the network type, number of nodes, number of objects, and network type relevant
parameters to draw an ER or PL network. We call this a configuration $\mathcal{P}$.
In addition, we fix the variance $\sigma$ of the anticipation distributions
$f_{i}$. We perform each simulation on $50$ different networks belonging to the
same configuration $\mathcal{P}$ and on each network we simulate the dynamics $50$
times. Then we average the obtained attendance distributions over all $2500$ simulations.

\paragraph{Datasets} To show the validity of our model we use real world recommender
datasets. {\bf MovieLens} (movielens.umn.edu), a  web service from GroupLens (grouplens.org) where ratings are recorded on a five stars scale. The data set contains $1682$ movies and $943$ users. Only $6,5\%$ of possible votes are expressed. {\bf Netflix} data set (netflix.com). We use the Netflix grand prize data set which contains $480189$ users and $17770$ movies and also uses a five stars scale. {\bf Lastfm} data set (Lastfm.com). This data set contains social networking, tagging, and music artist listening information from users of the Last.fm online music system. There are $1892$ users, $17632$ artists, and $92834$ user-listended artists relations in total. In addition, the
data set contains  $12717$ bi-directional user friendship relations. These data sets are chosen because they exhibit very different attendance distributions and thus provide an excellent playground to validate our model in different settings.

\paragraph{Experiments}
{\bf Data topologies}. We firstly investigate the simulated attendance distributions as a
function of trust $\gamma$, the anticipation threshold $\Delta$, and the network
topology. For this purpose we simulate the dynamics on a toy network with $500$ nodes and
record the final attendance number of $300$ objects. The simulation is conducted for
ER and PL networks and performed as outlined in the simulations paragraph above. In Fig.(\ref{fig:skew_heatmap_er}) 
and Fig.(\ref{fig:skew_heatmap_pw}) we investigate the skewness \cite{zwi00} of the attendance distributions and the maximal attendance
obtained for the corresponding parameter settings. The skewness of a distribution is
a measure for the asymmetry around its mean value. A positive skewness value means that there is more
weight to the left from the mean, whereas a negative value indicates more weight in the right from the mean.

{\bf Fitting real data}. We explore the model's ability to fit real world recommendation attendance distributions found in the described data sets. For this purpose we fix for the Netflix data set a
network with $480189$ nodes and perform a simulation for $17770$ objects. In the MovieLens case we do the same 
for $943$ nodes and $1682$ objects and for the Lastfm data set we simulate on a network with $1892$ 
nodes and $17632$ objects. In the case of Lastfm we have the social network data as well. We validate our model on that data set by two experiments: a) we use the provided user friendship network as simulation input and fit the attendance distribution and b) we fit the attendance distribution like in the MovieLens and Netflix case with an artificially generated network.

{\bf Mathematical analysis}. We investigate the Master Equations Eq.~(\ref{eq:mas1})
and Eq.~(\ref{eq:mas2}). We provide a full analytical solution
for Eq.~(\ref{eq:mas2}) and an analytical approximation for Eq.~(\ref{eq:mas1}) in the
early spreading stage. 

\section{Results }
\label{sec:results}
\begin{figure}[ht!]
\centering
\includegraphics[scale=0.35]{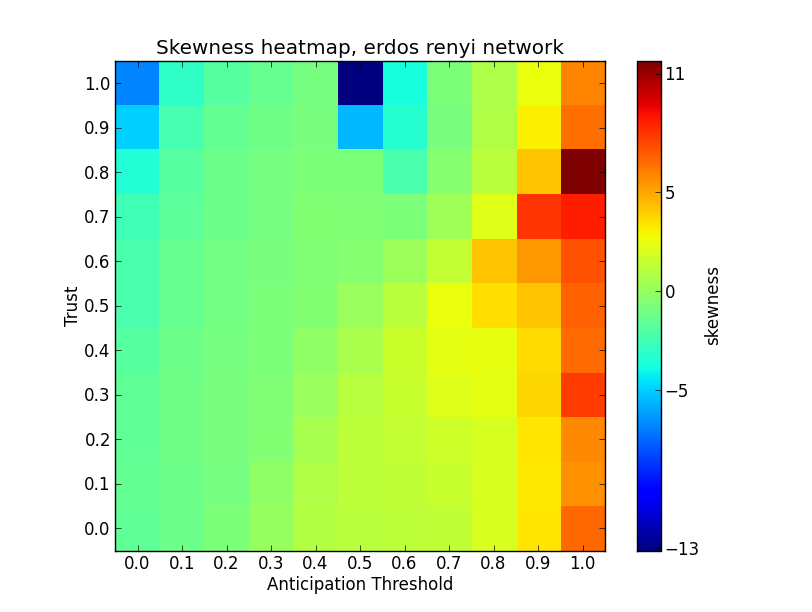}
\caption{Skewness of the attendance distributions as a function of trust $\gamma$ and the critical anticipation threshold $\Delta$ for Erd\H{o}s-R\'eny networks with $500$ nodes and $300$ simulated items.}
\label{fig:skew_heatmap_er}
\end{figure}

\begin{figure}[ht!]
\centering
\includegraphics[scale=0.35]{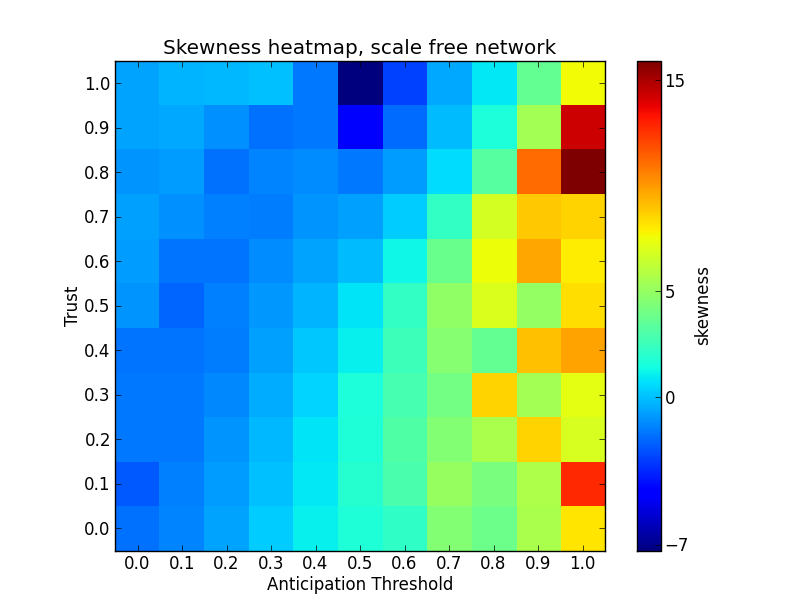}
\caption{Skewness of the attendance distributions as a function of trust $\gamma$ and the critical anticipation threshold $\Delta$ for power-law networks with $500$ nodes and $300$ simulated items.}
\label{fig:skew_heatmap_pw}
\end{figure}

{\bf Data topologies}. The landscape of attendance distributions of our model is demonstrated in Fig.~(\ref{fig:skew_heatmap_er}) and Fig.~(\ref{fig:skew_heatmap_pw}).  To obtain these results, simulations were performed as described in Sec.~(\ref{sec:methods}). The item anticipation $f_{i}$ was drawn from a normal distribution with mean values $\mu_{i} \in U(-0.1,0.1)$ and variance $\sigma=0.25$ fixed for all items. Both networks have $500$ nodes. In the Erd\H{o}s-R\'enyi case, we used a wiring probability $p=0.03$ between nodes. The Power Law network was drawn with an exponent $\delta = 2.25$.
The simulated attendance distributions in Fig.(\ref{fig:skew_heatmap_er}) 
and Fig.(\ref{fig:skew_heatmap_pw}) show a wide range of different patterns for both ER and PL Influence-Networks. In particular, both network types can serve as a basis for attendance distributions with both positive and negative skewness. Therefore, the observed fat-tailed distributions are not a result of the heterogeneity of a scale free network but they are emergent properties of the dynamics produced by our model. The parameter region for highly positively-skewed distributions is the same for both network types. The parameters $\gamma$ and $\Delta$ can be tuned so that all items are attended by everybody or all items are attended by nobody. While not relevant for simulating realistic attendance distributions, these extreme cases help to understand the model's flexibility.

\begin{figure}[ht!]
\centering
\includegraphics[scale=0.35]{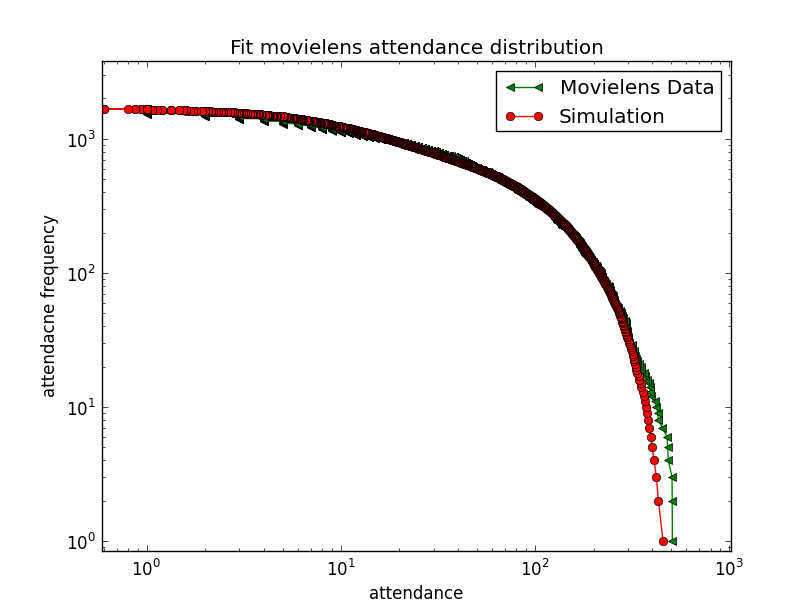}
\caption{Fit of the MovieLens attendance distribution with trust $\gamma = 0.50$, critical anticipation threshold $\Delta = 0.6$, anticipation distribution variance $\sigma=0.25$, and power law network with exponent $\delta=2.25$, $943$ nodes, and $1682$ simulated objects.}
\label{fig:fit_movielens}
\end{figure}

\begin{figure}[ht!]
\centering
\includegraphics[scale=0.35]{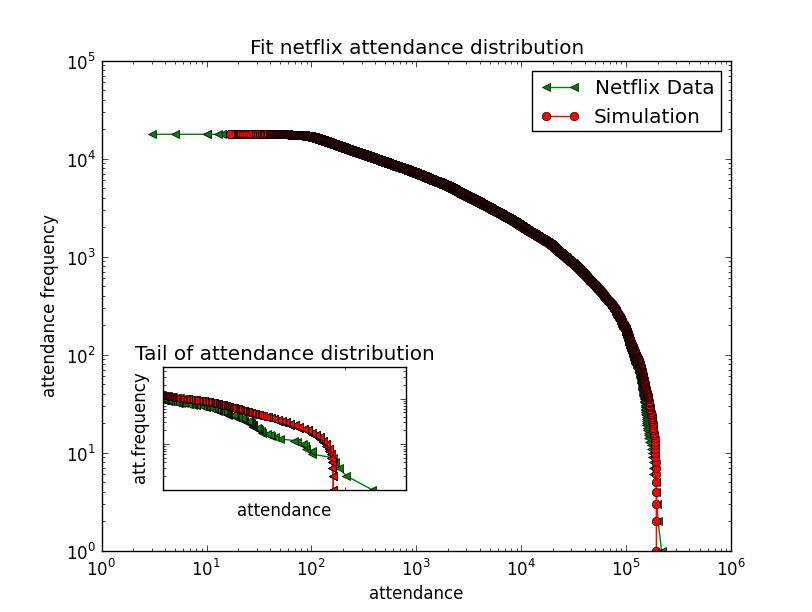}
\caption{Fit of the Netflix attendance distribution with trust $\gamma=0.52$, critical anticipation threshold $\Delta = 0.72$, anticipation distribution variance $\sigma=0.27$, and power law network with exponent $\delta=2.2$, $480189$ nodes, and $17770$ simulated objects.}
\label{fig:fit_netflix}
\end{figure}

\begin{figure}[ht!]
\centering
\includegraphics[scale=0.35]{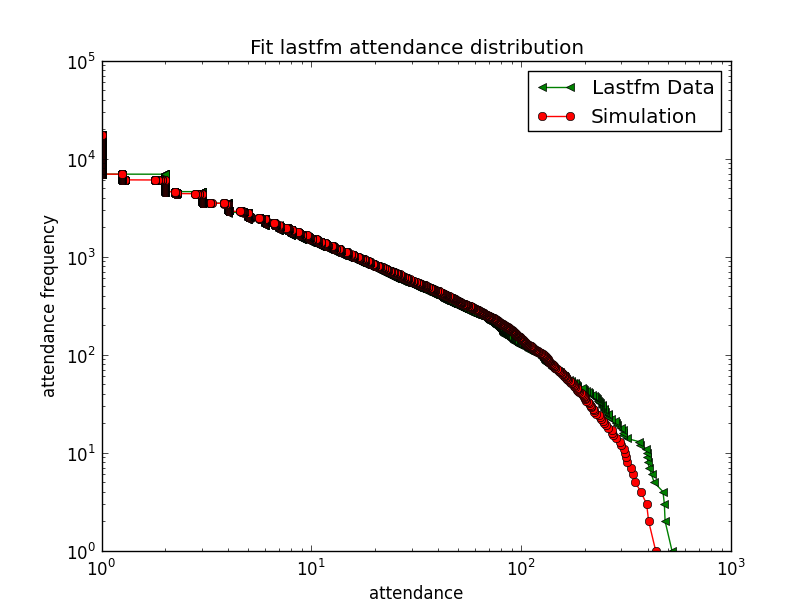}
\caption{Fit of the Lastfm attendance distribution with trust $\gamma = 0.4$, critical anticipation threshold $\Delta = 0.8$, anticipation distribution variance $\sigma = 0.24$, and real Lastfm user friendship network with $1892$ nodes and $17632$ simulated objects.}
\label{fig:fit_lastfm}
\end{figure}

\begin{figure}[ht!]
\centering
\includegraphics[scale=0.35]{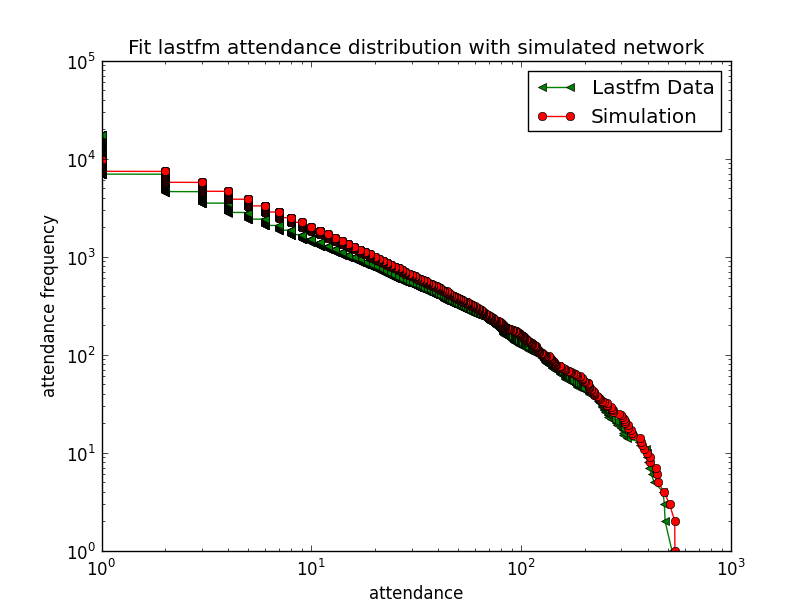}
\caption{Fit of the Lastfm attendance distribution with trust $\gamma = 0.6$, critical anticipation threshold $\Delta = 0.8$, anticipation distribution variance $\sigma = 0.24$, and power law network with exponent $\delta=2.25$, $1892$ nodes and $17632$ simulated objects.}
\label{fig:fit_lastfm_net}
\end{figure}

\begin{figure}[ht!]
\centering
\includegraphics[scale=0.35]{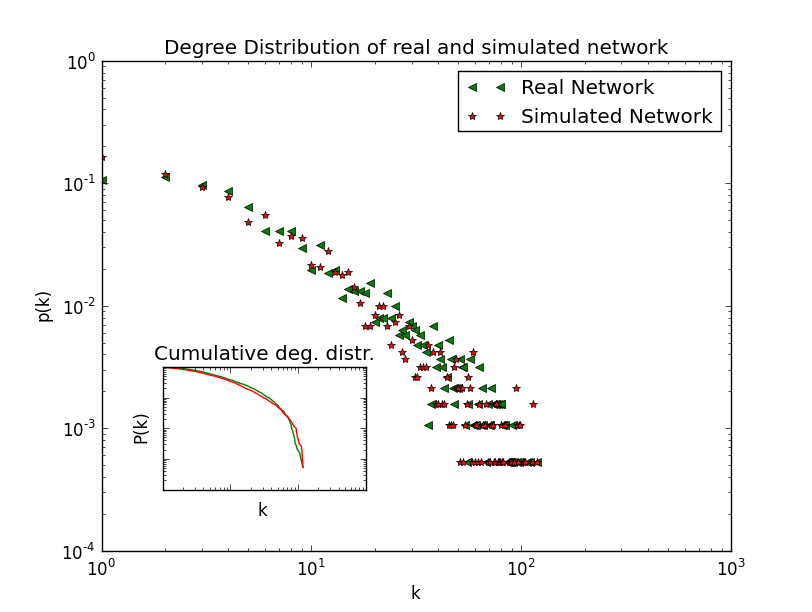}
\caption{Log-log plot of real (red) and simulated (blue) social network degree distribution $P(k)$ for the Lastfm data set. Inset: plot of the cumulative degree distribution.}
\label{fig:netLastfm}
\end{figure}

{\bf Fitting real data}
We fit real world recommender data from MovieLens, Netflix and Lastfm with results reported in Fig.~(\ref{fig:fit_movielens}), Fig.~(\ref{fig:fit_netflix}), Fig.~(\ref{fig:fit_lastfm}), Fig.~(\ref{fig:fit_lastfm_net}), and Tab.~(\ref{tbl:results1}), respectively. The real and simulated distributions are compared using Kullback-Leibler (KL) divergence \cite{kull51}. We report the mean,
median, maximum, and minimum of the simulated and real attendance distributions. Trust $\gamma$, anticipation threshold $\Delta$, and anticipation distribution variance $\sigma$ are reported in figure captions. We also compare the averaged mean degree, maximum degree, minimum degree, and clustering coefficient of the real Lastfm social network and networks obtained to fit the data. Results are reported in Tab.~(\ref{tbl:Lastfmres}) and Fig.~(\ref{fig:netLastfm}).
Note that thus obtained parameter values can be useful also in real applications where, assuming that our social opinion formation model is valid, one could detect decline of the overall trust value in an online community, for example.


\begin{table}
\begin{tabular}{l l c c c c}
D & KL & Med & Mean & Max & Min \\
\hline
ML & $0.046$ & $27/26$ & $59/60$ & $583/485$ & $1/1$ \\
NF & $0.030$ & $561/561$ & $5654/5837$ &  $232944/193424$ & $3/16$ \\  
LFM1 & $0.05$ & $1/1$ & $5.3/5.2$ & $611/503$ & $1/1$ \\ 
LFM2 & $0.028$& $1/1$ & $5.3/5.8$ & $611/547$& $1/1$ \\
\end{tabular}
\caption{Simulation results. ML: Movielens, NF: Netflix, LFM1: Lastfm with real
network, LFM2: Lastfm with simulated network, KL: Kullback-Leibler divergence, Med: Median, Mean, Max: maximal attendance (data/simulated), Min: minimal attendance (data/simulated).}
\label{tbl:results1}
\end{table}
\begin{table}
\begin{tabular}{l c c c c c}
D & $\mdeg{k}$  & $k_{min}$ & $k_{max}$ & $\delta$ & $C$\\
\hline
LFM1 & $13.4$ & $1$ & $119$ & $2.3$ & $0.186$  \\
LFM2 & $12.0$ & $1$ & $118$ &  $2.25$ & $0.06$ \\
\end{tabular}
\caption{Mean, minimum, maximum degree, clustering coefficient $C$, and estimated exponent $\delta$ of the real (LFM1) and simulated (LFM2) social network for the Lastfm data set.}
\label{tbl:Lastfmres}
\end{table}

{\bf Mathematical analysis}. Eq.~(\ref{eq:mas2}) can be solved analytically. We have $\forall t:\ a(t) + s(t) + d(t) = 1$ with the initial conditions for the first movers $a_{0}=
\int_{\Delta}^{u}f(x)dx$, $s(0) = 1 - a(0)$, and $d(0)=0$. In the following we use the bra-ket notation $\mdeg{x}$ to represent the average of a quantity $x$. Standard methods can now be used to arrive at\footnote{We give here only the solution for $a(t)$ because we are mainly interested in the attendance dynamics.}
\begin{equation}
a(t) =\frac{(\tau \left<k\right>)^{-1} \exp(t/\tau) }{(\alpha + \lambda)\left[\exp(t/\tau) - 1 \right] + (\tau \left<k\right>a_{0})^{-1}}.
\label{eq:an1}
\end{equation}
Here $\tau$ is the time scale of the propagation which is defined as
\begin{equation}
\tau = \left(a_{0} \alpha \left < k \right > + \lambda \left < k \right > \right)^{-1}.
\end{equation}
This is similar to the time scale $\tau= \left( \lambda \left <
k \right > \right)^{-1}$ in the well known SI Model \cite{new02,bar08}. Eq.(\ref{eq:an1})
can be very useful in predicting the average behavior of users in a recommender system. 

Since Eq.~(\ref{eq:mas1}) is not accessible to a full analytical solution, we investigate it for the 
early stage of the dynamics. Assuming $a(0)=a_{0} \gg 0$, we can neglect the dynamics of $d(t)$ to obtain
\begin{displaymath}
\dot{\Omega}(t) = \left( \frac{\mdeg{k^{2}}}{\mdeg{k}} - 1 \right) \Omega(t).
\end{displaymath}
In addition, Eq.~(\ref{eq:mas1}) yields
\begin{equation}
\left.\begin{aligned}
\dot{a}_{k}(t) &= \lambda k (1-a_{k}(t)) \Omega(t) \\
\dot{s}_{k}(t) &= -(\alpha + \lambda) k (1-a_{k}(t)) \Omega(t) \\
\end{aligned}
\right\}
\label{eq:mas3}
\end{equation}
Neglecting terms of order $a_{k}^{2}(t)$ and summing the solution of $a_{k}(t)$ over
$P(k)$, we get a result for the early spreading stage
\begin{equation}
a(t) = a(0) \Big(1 + \tau\lambda \big(\exp(t/\tau)-1\big) \Big),
\label{eq:masapprox}
\end{equation}
with the timescale $\tau = \mdeg{k^{2}}/\left[\lambda(\mdeg{k^{2}} - \mdeg{k})\right]$.
The obtained time scale $\tau$ valid in the early stage of the opinion spreading is
clearly dominated by the network heterogeneity. This result is in line with
known disease models, e.g., SI,SIR \cite{new02,bar08}. We emphasize that Eq.(\ref{eq:masapprox}) is valuable in predicting users' behavior of a recommender system in an early stage. 

\section{Discussion}
\label{sec:discussion}
Social influence and our peers are known to form and influence many of
our opinions and, ultimately, decisions. We propose here a simple model
which is based on heterogeneous agent expectations, a social network,
and a formalized social influence mechanism. We analyze the model by
numerical simulations and by master equation approach which is particularly
suitable to describe the initial phase of the social ``contagion''. The
proposed model is able to generate a wide range of different attendance
distributions, including those observed in popular real systems (Netflix,
Lastfm, and Movielens). In addition, we showed that these patterns are
emergent properties of the dynamics and not imposed by topology of the
underlying social network. Of particular interest is the case of Lastfm
where the underlying social network is known. Calibrating the observed
attendance distribution against the model then leads not only to social
influence parameters but also to the degree distribution of the social
network which agrees with that of the true social network.

The Kullback-Leibler distances (KL) for the simulated and real attendance
distributions are below $0.05$ in all cases, thus demonstrating a good fit.
However, the maximum attendances could not be reproduced exactly by the model.
One reason may be missing degree correlations in the simulated networks in
contrast to real networks where positive degree correlations (so-called degree
assortativity) are common. For the Lastfm user friendship network we observe
a higher clustering coefficient $C \approx 0.18$ compared to the clustering
coefficient $C \approx 0.06$ in the simulated network. To compensate for this,
a higher trust parameter $\gamma$ is needed to fit the real Lastfm attendance
distribution with simulated networks.

We are aware that our statistics to validate the model is not complete. But we are
confident, that our approach points to a fruitful research direction to understand recommender
systems' data as a social driven process.   

The proposed model can be a first step towards a data generator to simulate
bipartite user-object data with real-world data properties. This could be
used to test and validate new recommender algorithms and methods. Future
research directions may expand the proposed model to generate ratings within a
predefined scale. Moreover, it could be very interesting to investigate the
model in the scope of social imitation $\cite{mich05}$. 
\bibliographystyle{abbrv}
\bibliography{refer}
\end{document}